# Laser-written scalable sapphire integrated photonics platform

MOHAN WANG,[1,2] PATRICK S. SALTER,[1] FRANK P. PAYNE,[1] TONGYU LIU,[1] MARTIN J. BOOTH,[1] AND JULIAN A. J. FELLS[1,3]

[1]*Department of Engineering Science, University of Oxford, Parks Road, Oxford, OX1 3PJ, UK*
[2]*mohan.wang@eng.ox.ac.uk*
[3]*julian.fells@eng.ox.ac.uk*

**Abstract:** In this paper, we demonstrate the integration of photonic devices on sapphire substrates using multi-layer depressed cladding waveguides at both 780 nm and 1550 nm. The devices are up to 10-cm long and written at depths down to 400 µm. The propagation losses for single-mode guiding are ~ 0.6 dB/cm at 780 nm and ~ 0.7 dB/cm at 1550 nm. A number of structures have been fabricated with simultaneous single-mode and polarization independent operation: evanescently coupled waveguide arrays, Y-branch splitters, Mach-Zehnder interferometers, and a 2×2 directional-coupler. All the devices were fabricated using adaptive optics-assisted femtosecond laser direct writing with a customized laser writing algorithm. This work enables the integration of single-mode sapphire photonics devices in a scalable manner, enabling many applications in communications, imaging, computing, and sensing.

## 1. Introduction

Integrated photonic devices are the optical equivalent of electronic integrated circuits. The fabrication process involves combining various photonics devices on a single substrate that are useful for generating, focusing, splitting, combining, isolating, polarizing, coupling, switching, modulating and detecting light [1]. Since the first demonstration by Davis *et al*. in 1995 [2], femtosecond laser fabrication has been widely established for fast prototyping of integrated photonics devices. By tightly focusing the high intensity laser pulse in a transparent substrate and moving the sample along a pre-programmed trajectory, customized 3-D structures can be written [3].

Crystals like sapphire, have excellent properties, including hardness, a wide spectral transmission window, and very good environmental stability, making them attractive photonics platforms for applications such as lasing, quantum computing, and sensing [4]. There are three main types of femtosecond laser-written waveguides in crystals: Firstly, waveguides formed by a laser-induced refractive index increase at low fluence [5]. These are similar to glass waveguides and are easy to construct, but susceptible to thermal decay [6]. Secondly, waveguides utilizing the stress-induced refractive index increase between two neighboring tracks, laser-written at a higher fluence. The drawbacks of these waveguides include their polarization-selective guiding and inability to support longer wavelength designs [7], which limit their application.

Thirdly, depressed cladding waveguides (DCWs) can be created by laser-writing an inner cladding of a finite diameter and a lower refractive index, with the core left undisturbed. The geometry of the laser-written inner cladding can be flexibly customized to be rectangular [8], circular [9], or lattice-like [10–13]. Depressed cladding waveguides are extremely environmentally stable and polarization insensitive. The first femtosecond laser-written DCW in crystal, a rectangular waveguide in Nd:YAG, was reported by Okhrimchuk *et al*. in 2005 [8]. So far, single- and multi-mode laser-written DCWs have been demonstrated in a wide variety of crystals, including Nd:YAG [8,14], Yb:YAG [15], LiNbO$_3$ [9,10], Ti:sapphire [12,16], silicon [17], and single-crystal sapphire [11,13,18,19,20,21,22], and for a range of different wavelengths.

Driven by the above-mentioned advantages, extensive research over the past decade has focused on developing femtosecond laser-written photonic devices based on DCWs for applications in lasing and sensing. Y-branch splitters have been demonstrated using a square-shaped cladding in Ti:sapphire [23] and MgO:LiTaO$_3$ [24]; using circular-shaped cladding in LiNbO$_3$ [25,26], Tm:KluW [27], and Nd:YAG [28]; and using lattice-like cladding in LiNbO$_3$ [10], sapphire [11,13], and Ti:sapphire [12]. By putting two Y-splitters back-to-back, a Mach-Zehnder interferometer (MZI) has been demonstrated [25]. Directional couplers have been reported in LiNbO$_3$ [29] and Tm:YAG [30], with a near perfect splitting ratio in the later. Meanwhile, light propagation in a weakly guiding and evanescently coupled photonics waveguide array in glass has been widely used as a model to mimic quantum systems due to its discreteness [31]. An evanescently coupled waveguide array in crystal could enable further exploration of optical effects that are related to the crystal properties. The first waveguide array was demonstrated by Heinrich *et al.* in 2018 using a stress-waveguide in LiNbO$_3$ [32]. Then more recently by Liu *et al.* using a stress-waveguide in Nd:YAG [33]. However, all the above-mentioned works rely on single-layer depressed cladding design, leaving the flexibility of geometry unexplored and large room for optimization.

One challenge to fabricating a DCW is the asymmetric shape of the laser modification, an intrinsic and undesirable characteristic of transversal laser writing [34]. To address this, extra layers are added in the horizontal direction of a DCW [15], compensating for the asymmetric laser modification to improve the guiding performance. Another challenge is the intrinsic leakage loss in a DCW due to the limited cladding size, as predicted from a dual cladding fiber model [35]. In our previous work, we have demonstrated experimentally that the propagation loss can be significantly reduced by increasing the cladding-to-core diameter ratio [36]. However, the cladding geometry can become increasingly complicated. In our previous work, a single-mode waveguide at 1550 nm with a loss of 0.8 dB/cm had 134 tracks [36], while a single-mode waveguide designed for the MIR wavelength range with a loss of ~ 0.4 dB/cm needed as many as 705 tracks [20,21]. Constructing devices based on DCW with complex geometry has, so far, limited the integration of DCWs into scalable photonics devices.

In this work, we extend the state-of-the-art in sapphire waveguide technology into more complex structures with the potential for a wide range of devices, building on our preliminary results [37]. The fabrication and characterization are described in Section 2. Sections 3, 4, 5, and 6 show the fabrication of a Y-branch splitter, an evanescently coupled waveguide array, a directional coupler, and an MZI, respectively. Devices with smaller dimensions for operation at 780 nm are demonstrated in Section 7. We believe these new devices will have many wide and diverse applications.

## 2. Fabrication and characterization

Figure 1(a) shows an illustration of the fabrication system. A regenerative femtosecond laser (Pharos SP06-1000-PP), delivering femtosecond laser pulses at a second harmonic generation wavelength of 515 nm and a pulse duration of 170 fs, was used for the fabrication. The pulse energy was adjusted externally by tuning a motorized half-wave plate placed in front of a polarizing beam splitter, while the power was constantly monitored using an external photodetector. A pulse energy of 30 nJ was used to fabricate all the 1550-nm waveguides. The laser beam was first expanded using a telescope set up to be projected onto a spatial light modulator (SLM, Hamamatsu X10468), then focused onto the pupil plane of a 40× objective (Zeiss EC Plan-Neofluar, NA 0.75).

Sapphire substrates (PI-KEM Inc.) with dimensions of 10 × 10 × 1 mm, 40 × 10 × 1 mm, and 100 × 10 × 1 mm were used in this work, with the *c*-axes along the waveguide direction. The samples were mounted on a three-axis nano-precision motion stage (Aerotech ABL10100L for *x* and *z* axes and ANT95-3-V for the *y* axis). The laser beam was linearly polarized, with the polarization direction aligned with the sapphire optical axis. Devices were written with a constant 100,000 pulses per mm or a pulse-to-pulse overlap of 10 nm, using a variable speed of between 1 and 10 mm/s and a repetition rate of between 100 kHz and 1 MHz. The center of the devices was 400 µm below the substrate surface. During fabrication, a spherical aberration correction phase pattern was adaptively applied to the SLM, calculated using position feedback provided by the stage [38]. The phase pattern for correcting spherical aberration for a depth of 400 µm is shown on the SLM in Fig. 1(a). After fabrication, the chips were mounted on a six-axis mechanical stage. Light from a tunable laser source (Agilent 8164A) was butt-coupled into the waveguide from a polarization maintaining single-mode fiber (Thorlabs P3-1310-PM) fixed in a rotary mount.

Figure 1(b) shows the design of the 1550-nm DCW. Each red ellipse represents a laser written track, in the form of a horizontal scan across the full length of the substrate. The waveguide has a core diameter of 10 µm and a square shaped cladding with a side length of ~ 40 µm. The number of tracks is 88, which is less than two thirds of our previous design. Although the tracks were not closely overlapped, they were found to be sufficiently effective for low-loss and polarization-independent guiding. Figure 1(c) shows a microscope image of the cross-section of the fabricated DCW, written using a speed of 10 mm/s and a repetition rate of 1 MHz.

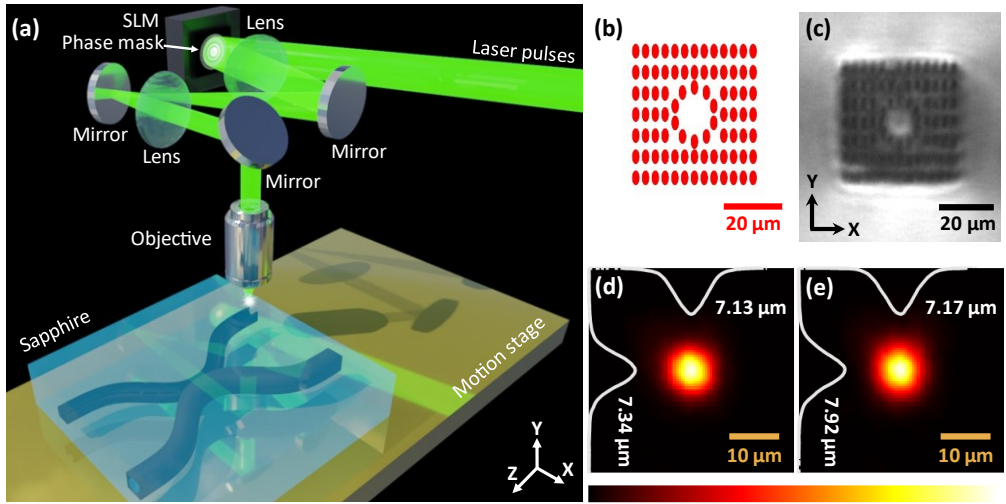

**Fig. 1.** (a) Schematic illustration of the femtosecond laser fabrication setup. The aberration correction phase pattern is displayed on the SLM; (b) the design of a single-mode waveguide for the 1550 nm wavelength, with each red ellipse representing a laser-written track; (c) a microscope image of the waveguide cross-section; (d-e) the near-field mode profiles of a DCW written on a 100 × 10 × 1 mm sapphire substrate for TE(d) and TM(e) polarizations, with the intensity profiles and FWHM for the horizontal and vertical directions plotted and labeled on the top and left edges. Experimental results for 1550 nm are displayed using the *hot* colormap. A color bar is shown beneath (d) and (e).

Figure 1(d-e) shows the near-field mode profiles of a 100-mm long DCW for the TE and TM polarizations, respectively. The full-width-half-maximum (FWHM) diameters are labeled on the respective images. No significant polarization dependence was observed. By comparing the experimentally measured FWHM with simulations in FIMMWAVE (Photon Design Ltd.), the refractive index change was calculated to be $\Delta n \approx -3.1 \times 10^{-3}$ at 1550 nm ($n_{sapphire}$ = 1.74628 at 1550 nm). This yielded a *V*-number of 2.11, confirming that our waveguide is single-mode.

The loss measurement was performed using a power meter (Newport 1919-R power meter and 818-IG detector). The propagation loss was measured from the difference between the total insertion loss from three waveguides written on 10-mm, 40-mm, and 100-mm substrates. By averaging several sets of measurements from waveguides written on the substates of the three different lengths, the propagation loss was determined to be ~ 0.7 dB/cm for both the TE and TM polarizations. The leaky mode loss predicted from FIMMWAVE simulation was 0.015 dB/cm, which was much lower than the experimentally measured loss. This suggested the existence of other significant loss mechanisms arising from the fabrication. One explanation for the excessive loss was scattering loss from the generation of sub-wavelength-scale nano-scatterers inside the depressed cladding [39].

Figure 2 shows the customizable software block diagram to generate the motion stage code for device fabrication. The device can have an arbitrary number of waveguides that can have non-parallel and overlapping trajectories. The algorithm is divided into four main blocks: 1) the construction of the waveguide cross-section, which is important to optimize since the waveguide quality directly determines the quality of the resulting devices. This optimization is described in our previous work [36]. The output is a 2-D array of track positions in the *X-Y* plane; 2) the design of a device from simulation, where 2-D arrays for the trajectory of each waveguide in the *X-Z* plane are defined; 3) coordinates in 3-D are generated for each track within each waveguide at varying step sizes, along the laser writing direction. Predefined finer resolutions are applied to the sections that involve either bending or waveguide overlap. A filter is then programmed to detect any overlap between neighboring waveguides for each step along writing direction, creating a flag to avoid repeated writing; and 4) the code output to the high-precision motion stage is generated. During fabrication, the laser repetition rate is changed adaptively with the writing speed using the stage PSO control [36] to create a constant 10,000 pulses per mm.

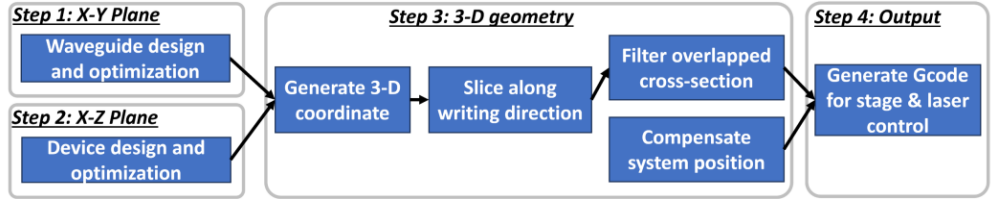

**Fig. 2.** The software block diagram for the writing of scalable photonics devices.

## 3. Y-branch splitter

Depressed cladding waveguides with different S-shaped bend designs were written using a pulse energy of 30 nJ, a speed of between 5 and 10 mm/s and a repetition rate of between 500 kHz and 1 MHz. The top view of five waveguides with different bending curvatures can be seen in Fig. 3(a). The shape of the bending section was designed as [40]:

$$x(z) = \frac{zS}{L} - \frac{S}{2\pi}\sin\left(\frac{2\pi z}{L}\right) \qquad (1)$$

The vertical distance, *S,* was set to 125 μm. Waveguides were fabricated for a range of different horizontal distances, *L,* of 3, 4, 5, 6, and 7 mm, shown respectively from top to bottom in Fig. 3(a). The separation between DCWs was 100 μm, which was sufficient to ensure no coupling between neighboring waveguides.

Figure 3(b) shows a magnified 1:1 ratio photo of the *L* = 3 mm waveguide at the bending region. The step size used for the bending region of 50 μm can be clearly identified from the top view, due to the imperfect speed profile. Acceleration and deceleration took place at the beginning and end of each step, causing inhomogeneities in the refractive index change along the light propagation direction. This is likely the reason for the discontinuous cracks, visible in Fig. 3(a), at certain positions due to the material releasing the accumulated stress. However, the influence of the cracks was minimal as they only happened at the periphery of the waveguide (usually at the top) and did not impact the guiding performance.

The loss due to bending was measured to be the difference in the total insertion loss between that of a bent waveguide and that of a neighboring straight waveguide. The table in Fig. 3(c) summarizes the measurement results for both polarizations. While the bend radius varies with the distance in Eq. (1), the maximum bending curvatures calculated using $r_{max} = L^2/2\pi S$ are listed in Fig. 3(c). The large experimentally measured loss is very likely caused by fabrication defects. Future work is needed to optimize the speed profile of the software, in order to maintain a constant speed for any given 3-D design.

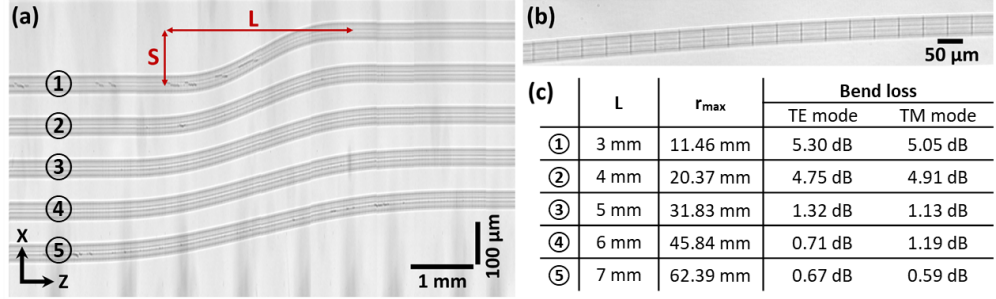

Fig. 3. (a) The top view of S-shaped bent single-mode waveguides with $S=125$ μm and a range of different $L$, of between 3 and 7 mm. All the top-view photos in this manuscript except (b) were made by stitching multiple images together, due to the limited field of view of our microscope camera, (b) the magnified view of bending region of the topmost waveguide in (a), showing fabrication defects at the start and end of each pre-defined step along the writing direction, which was due to acceleration and deceleration from an imperfect speed profile, and (c) the measured bend losses of the different designs in (a).

A Y-splitter with $L=5$ mm and $S=125$ μm, yielding a divergence angle of ~ 2.89°, was written on a $10 \times 10 \times 1$ mm$^3$ sapphire chip. The two output ports of the splitter were 250 μm apart, to match the separation of a V-shaped fiber groove. A pulse energy of 30 nJ was used. The speed and repetition rate were adaptively adjusted between 10 mm/s and 1 MHz for the straight region and 1.25 mm/s and 125 kHz for the S-shaped bend region. Simulations of the devices were performed using the software package FIMMWAVE assuming a homogenous and infinitely large cladding. The top view of the simulation and the laser-written device are shown side-by-side in Fig. 4(a) and (b). A photo of the output cross-section is shown in Fig. 4(c). The near-field mode profiles for both polarizations (measured as described previously) are shown in Fig. 4(d) and (e). The splitting ratio was found to be 47:53 and 50:50 for the TE and TM polarizations, respectively.

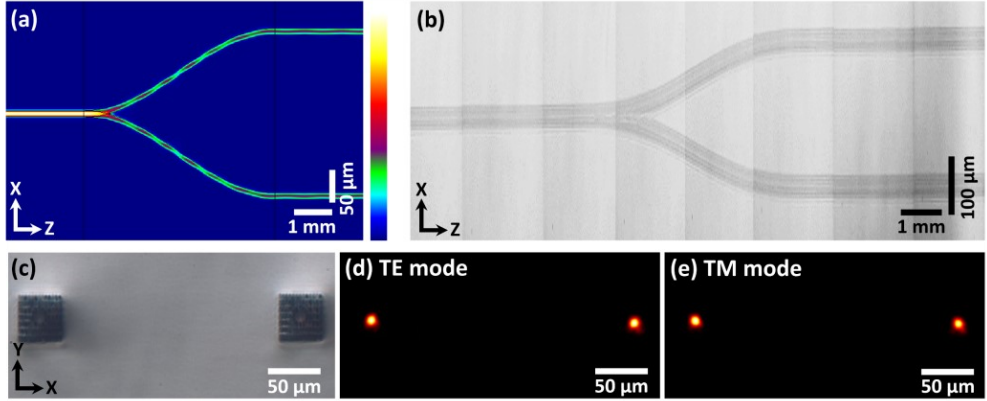

**Fig. 4.** (a) Simulation result of a Y-branch splitter. Simulation results in this manuscript are displayed using the FIMMWAVE *thermal-plus* colormap. A color bar is shown on the right, (b) the microscopic top view of a laser-written Y-shaped splitter, (c) the microscopic cross-section of a Y-splitter, and (d-e) the near-field mode profiles of the Y-splitter output at 1550 nm for TE (d) and TM (e) polarizations.

## 4.  Evanescently coupled waveguide array

A waveguide array can be used to evaluate the coupling behavior, for precise construction of evanescently coupled devices such as directional couplers, or as an experimental platform for simulating solid-state physics structures [41]. An array of seven evanescently coupled waveguides was designed to have a center-to-center separation of 13.1 µm, allowing one track between two adjacent waveguides. Waveguide arrays with interaction lengths of 3, 4, and 5 mm were written on a 10 × 10 × 1 mm substrate, using a speed of 10 mm/s, a repetition rate of 1 MHz, and a pulse energy of 30 nJ. The different interaction lengths allowed a non-invasive observation of the diffraction pattern evolution along the laser propagation direction. Figure 5(a) shows the cross-section of a fabricated array. The top view of the three arrays is shown in Fig. 5(b), with the input on the left and output on the right.

The diffraction patterns for both polarizations were recorded using the imaging system. Figure 5(c-h) shows the near-field mode profiles, with the TE mode on the top and the TM mode on the bottom, for the three interaction lengths, respectively. The diffraction patterns were mostly symmetric, with only slight distortion of the mode field for the TE polarization. The coupling coefficient is also slightly higher for TE than TM polarization, when comparing Fig. 5(g) with (h). We found both to be a result of the stress introduced during fabrication, which can be mitigated from thermal treatment.

Simulations were performed in FIMMWAVE to match the experimentally observed diffraction patterns with the simulated results. However, the estimated refractive index change calculated from the near-field profile in Section 2 was found to be a poor fit, as the FWHM of the waveguide modes in Fig. 5(c-h) appeared to be smaller than that of a single waveguide in Fig. 1(d-e). A set of simulations with the refractive index change parameter swept at a step-size of $-0.1 \times 10^{-3}$ were performed, to find the laser-written cladding index decrease, using a technique similar to that described in [42]. A best fit was found to be $\Delta n = -4.0 \times 10^{-3}$, with the corresponding simulation result shown in Fig. 5(i). The diffraction patterns at the three interaction lengths, labeled with arrows, were compared with the measurements. A good match was found between the simulation and experimental results. The reason for the discrepancy in the refractive index change is not clear, though an assumption is the enlarged cladding could cause an accumulated increased refractive index change in the cladding. It would require further study to profile the refractive index distribution and to better understand the mechanism.

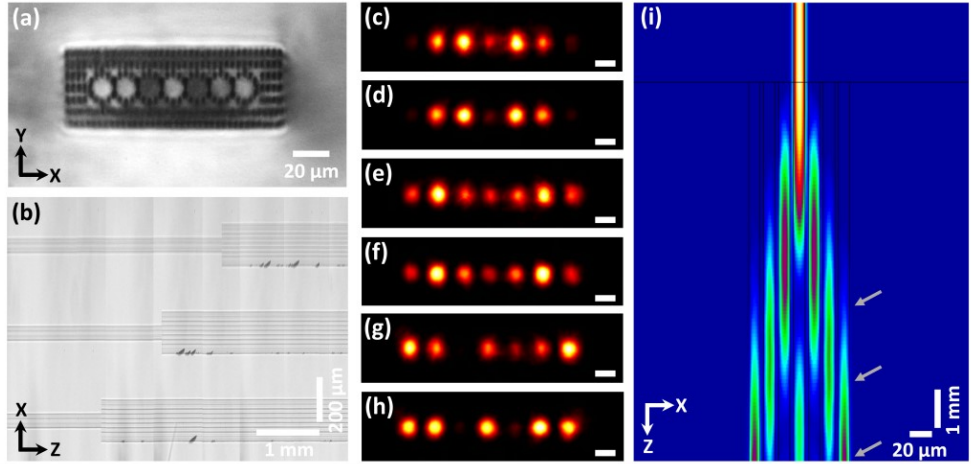

Fig. 5. (a) Microscope image of the cross-section of an evanescently coupled DCW array, composed of seven waveguides with a center-to-center separation of 13.1 µm, (b) the top-view of three such waveguide arrays with interaction distances of 3, 4, and 5 mm, respectively. Light was launched from the DCW on the left into the center waveguide of the array on the right, (c-h) the diffraction profiles of the waveguide arrays at 1550 nm for both polarizations, with an interaction distance of 3 mm (c-d), 4 mm (e-f), and 5 mm (g-h), with the TE mode on the top and TM mode on the bottom. The scale bars in (c-h) are 10 µm, (i) the simulation result for a 5-mm long waveguide array. The arrows point at interaction distances of 3, 4, and 5 mm.

## 5. Directional coupler

Although the characterization of the evanescently coupled array in Section 4 should provide sufficient information for a directional coupler design, daily laser power fluctuations could also lead to variations in the laser-written refractive index. A calibration, prior to writing the directional coupler, was performed on 1 × 2 linear arrays that have the same cross-section as if the directional coupler was cut in half, as shown in Fig. 6(a). The same core-to-core distance of 13.1 µm was chosen and different interaction lengths were fabricated. The linear arrays were written on a 10 × 10 × 1 mm substrate. Each array consisted of an input waveguide followed by the 1 × 2 waveguide array, with the lengths appropriately adjusted to provide a range of interaction lengths.

Light with a mixed polarization from a single-mode fiber (SMF28e+) was coupled into the input waveguide. Figure 6(b-f) shows the measured mode fields for light injected into the lefthand waveguide, with the respective interaction lengths shown on top of each figure. A coupling length of $L_c$ = 1.75 mm provides a 50:50 intensity splitting between the two waveguides, yielding a coupling constant of $c = \pi/2L_c \approx 0.90$ rad/mm. These results were compared with the simulation results, shown in Fig. 6(g), with arrows pointing to the positions of the interaction lengths of Fig. 6(b-f). The refractive index change was found to be $\Delta n = -3.15 \times 10^{-3}$, close to the estimate from the mode field in Section 2 and much smaller than that of the evanescently coupled array in Section 5.

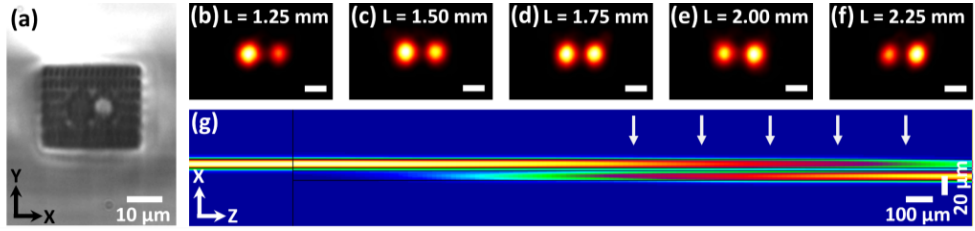

Fig. 6. (a) Microscope image of the cross-section of the interaction region of a directional coupler; the core-to-core separation is 13.1 µm; (b-f) the experimental mode field measurements from 1 × 2 waveguide arrays with interaction lengths from 1.25 to 2.25 mm with a step of 0.25 mm; light was injected into the waveguide on the left. The scale bars are all 10 µm, and (g) the simulated top view of the 1 × 2 linear array. The arrows point to the positions of the interaction lengths corresponding to (b-f).

Using this calibrated refractive index change, a 2 × 2 directional coupler was designed using simulations in FIMMWAVE for 50:50 beam splitting. The same S-shaped bend design from Section 3 was used and the output ports had a separation of 250 µm. The straight region in the middle was designed to be 840 µm. Figure 7(a) shows the top view of the simulation result. The directional coupler was fabricated using 30 nJ pulse energy, 1 to 10 mm/s speed, and 100 kHz to 1 MHz repetition rate on a 40 × 10 × 1 mm sapphire substrate. The top view of the fabricated device is shown in Fig. 7(b). The measured near-field mode profiles at 1550 nm for the TE and TM polarizations are shown in Fig. 7(c) and Fig. 7(d), respectively. The coupling ratio was measured to be 50:50 for both polarizations, with minimal polarization dependence.

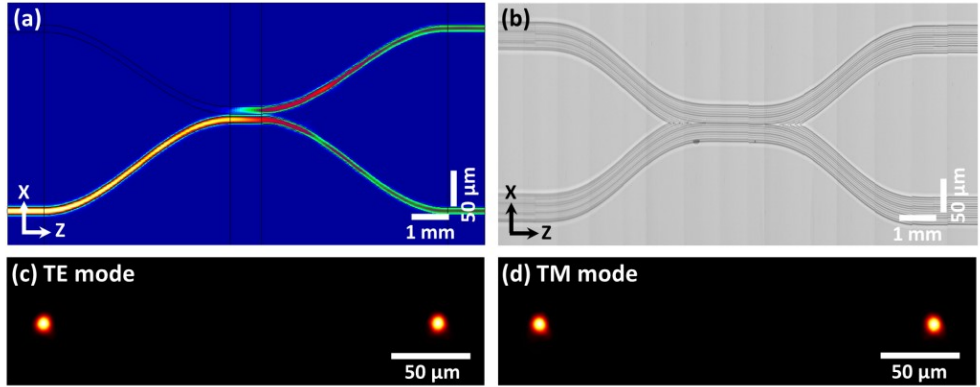

Fig. 7. (a-b) The top view of the simulation result (a) and fabricated device (b) for a directional coupler, with an input and output port spacing of 250 µm and interaction region length of 840 µm, (b) the top-view with a straight region, and (c-d) the near-field mode profiles of TE (c) and TM (d) polarizations at 1550 nm.

## 6. Mach-Zehnder Interferometer

Two Y-branches, with the same design as described in Section 3, were used to form a balanced MZI, by combining them back-to-back. The MZI was written on a 40 × 10 × 1 mm sapphire substrate. The distance between the input Y-branch splitter and the output Y-branch combiner was chosen to be 5 mm. The same pulse energy, repetition rate, and speed for the Y-branch splitters were used as before. Figure 8(a) shows the top-view fabricated device. A microscope image of the output facet is shown in Fig. 8(b), and the measured mode fields for the TE and

TM polarizations are shown in Fig. 8(c) and Fig. 8(d), respectively, demonstrating polarization insensitive performance.

Fig. 8(e) shows an unbalanced MZI with two arms of different lengths. The MZI was written using the same fabrication parameters as above. The asymmetric Y-branches were formed from S-bends with different parameters. The parameters for the top S-bend were $S$=1.5 mm and $L$=17.5 mm. The parameters for the bottom S-bend were $S$=0.125 mm and $L$=17.5 mm. The two arms had a length difference of $\Delta L$=190.4 µm. The transmission spectrum was measured using the same measurement set up described in Section 2.

The power meter and the tunable laser source were used to measure the transmitted power over a spectral range of 80 nm in steps of 0.25 nm, as shown in Fig. 8(f). The fringes are slightly distorted by the Fresnel reflection between the two sapphire facets. The theoretical free spectral range given by $\lambda^2/(\Delta L \cdot n_{sapphire})$ is approximately 7.23 nm, which matches well with the experimental measurement. The fringe visibility could be further improved by optimizing the bending loss. The MZI demonstrated here could potentially be combined with microfluidic channels created from laser micromachining techniques on sapphire [43], as an alternative and more chemically resistant platform for biosensing [44], or used in cascaded form for broadband spectral filters [45]. The same technology could also be applied to other crystalline platforms for high-quality optical switches.

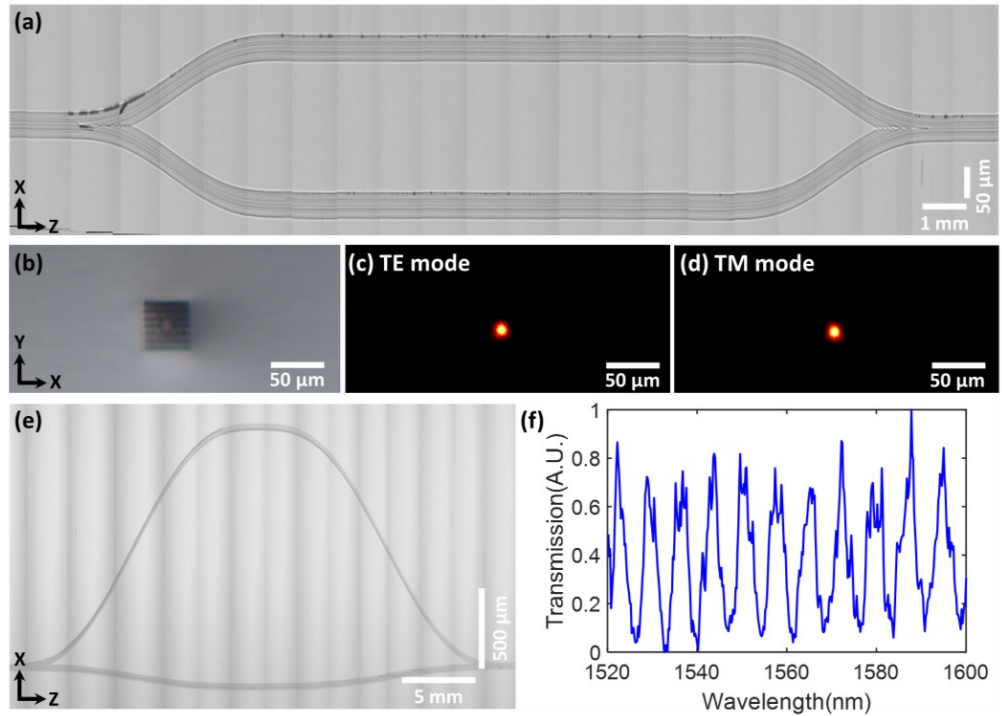

**Fig. 8** (a) The top view of an MZI with balanced arms, made from two Y-branch splitters connected back-to-back, (b-d) the microscopic cross-section (b), and the near-field mode profiles for TE (c) and TM (d) polarizations of the MZI output, (e) the top view of an unbalanced MZI. The two arms have a length difference of 190.4 µm, and (f) the transmission spectrum of the unbalanced MZI output. The spectrum was slightly distorted by fringes formed by the Fabry-Pérot cavity created from Fresnel reflections at the two sapphire facets. A Fourier transform of the fringe suggests a cavity length of 184.4 µm, which closely matches the design.

## 7. Operation at 780 nm

In this section, we demonstrate how the above-mentioned techniques can be used to write DCWs operating at 780 nm using smaller geometries. Initially, the refractive index change at 1550 nm was used for a coarse estimation for designing the waveguide for 780 nm. A set of DCWs with a core diameter from 5 µm to 8 µm were fabricated, all with a circular cladding diameter of ~ 40 µm. Although a circular-shape cladding was used here, a square-shape cladding with a well-defined design would provide very similar results.

The DCWs were fabricated using a lower pulse energy of 25 nJ, a repetition rate of 1 MHz, and a speed of 10 mm/s. After fabrication, the waveguides were measured as previously, using a 780 nm fiber-coupled single-mode light source (Thorlabs S1FC780) and a polarization maintaining fiber (Thorlabs P3-780PMY) which was single-mode at 780 nm. The launching condition was varied using a six-axis positioning stage. For core diameters of larger than 7 µm, higher-order modes ($LP_{11}$) could be excited at certain launching angles, indicating a few-mode waveguide. A core diameter of 6.5 µm was found to provide the most optimized single-mode guiding.

Figure 9(a) shows the optimized design for a single-mode waveguide at 780 nm, which was composed of 78 tracks. The waveguide cross-section is shown in Fig. 9(b). The near-field mode fields for TE and TM polarizations are shown in Fig. 9(c) and Fig. 9(d), respectively. The cladding refractive index change was estimated using the FWHM of the mode fields to be − 2.3 × $10^{-3}$ ($n_{sapphire}$ = 1.7607 at 780 nm). This gives a V number of 2.35, confirming single-mode operation. Devices with this design were fabricated on 10-mm, 40-mm, and 100-mm long sapphire substrates, at a depth of 400 µm below surface. The propagation loss measured from the difference between the total insertion loss of the waveguides at different lengths was found to be ~ 0.6 dB/cm for the TE polarization and ~ 0.5 dB/cm for the TM polarization.

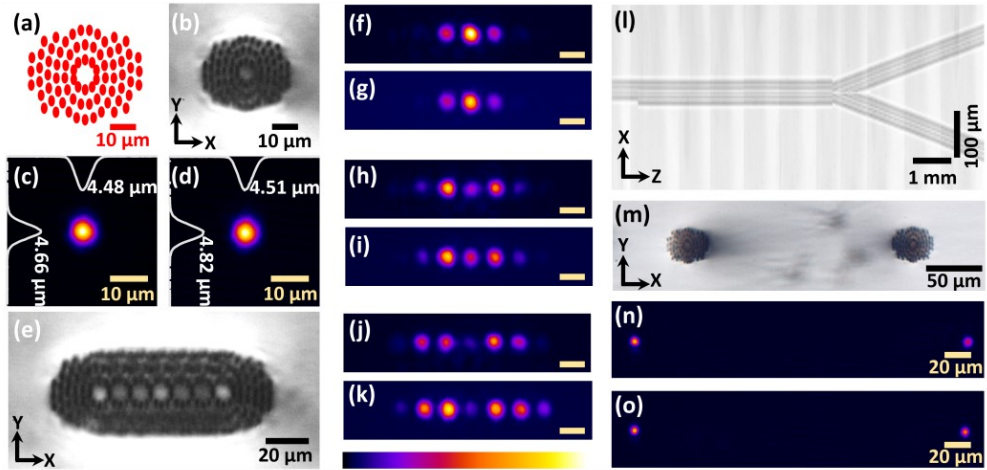

**Fig. 9**. (a-b) The design (a) and microscope image of the cross-section (b) of a single-mode DCW for 780 nm wavelength, (c-f) the near-field mode profiles at 780 nm for TE (c) and TM (d) polarizations, (e) the microscopic cross-section of a 1 x 7 horizontal array with a core-to-core separation of 9.8 µm, (f-k) the diffraction profile of the linear array for the two polarizations with an interaction distance of 30 mm (f-g), 50 mm (h-i), and 70 mm (j-k). The scale bars are 10 µm, (l-m); the top view (l) and cross-section (m) of a Y-branch directional coupler splitter with an interaction length of 5 mm, a straight splitter section of 4 mm, and an output core-to-core distance of 250 µm, and (n-o) the mode profile of the Y-branch splitter for TE (n) and TM (o) polarizations at 780 nm. The experimental results for 780 nm are displayed using the *fire* color map. A color bar is shown beneath (k).

Linear arrays with a core-to-core separation of 9.8 µm and interaction lengths of 3, 5, and 7 mm were fabricated on a 10-mm long sapphire chip, following similar designs as described in Section 4. A cross-sectional microscope image of a linear array is shown in Fig. 9(e). Figure 9(f-k) are the near-field mode profiles of both polarizations for the three different interaction lengths, with the TE polarization in the top and TM polarization in the bottom. The results show polarization independent performance at 780 nm.

A Y-branch directional coupler was fabricated for the 780 nm wavelength on a 10 × 10 × 1 mm sapphire substrate. The design is similar to our previous 1550 nm design reported in [37]. The interaction length of the directional coupler was designed to be 5 mm for 50:50 splitting. The top view and output cross-sectional view are shown in Fig. 6(l) and (m). Light from a polarization-maintaining fiber was coupled into a 1-mm long single-mode DCW which fed a 5-mm long directional coupler region with a core-to-core separation of 9.8 µm. Finally, the output from both directional coupler outputs were split via a splitting angle of 3.58° using a linear fit over a distance of 4 mm, into two output ports which were 250 µm apart. The measured mode field in TE and TM polarizations are shown in Fig. 6(m) and (n), respectively.

## 8. Conclusion

In this work, optical devices based on single-mode multi-layer DCWs were fabricated in sapphire substrates. Fabrication was performed using adaptive optics-assisted femtosecond laser direct writing. We demonstrated single-mode DCWs at both 780 nm and 1550 nm that are up to 10-cm long, with propagation losses of ~ 0.6 dB/cm and ~ 0.7 dB/cm, respectively. A Y-branch splitter and a 2 × 2 directional coupler were designed for the 1550 nm wavelength range. The measurement showed a near perfect splitting ratio. Evanescently coupled waveguide arrays were fabricated, with the experimental measurement in good agreement with the simulation results. Mach-Zehnder interferometers were designed by cascading a Y-branch splitter and a Y-branch combiner. The transmission spectrum was measured, showing the interference fringe separation to be consistent with the optical path difference between the two arms. The technique was then flexibility tailored to the 780 nm wavelength range by adjusting the waveguide dimensions. A waveguide array and a Y-branch splitter were fabricated at 780 nm. All the devices exhibited polarization independent performance.

The work has proven the concept of fabricating compact and scalable photonics on a single-crystal sapphire platform using direct laser writing. We believe the approach demonstrated here will open up exciting opportunities in the design and manufacturing of integrated photonic chips on a variety of crystalline platforms, for applications such lab-on-a-chip sensing, high-quality lasing, and quantum and optical computing.

**Funding.** Engineering and Physical Sciences Research Council (EP/T00326X/1).

**Acknowledgments.** The authors gratefully acknowledge the support and advice of their partners Rolls-Royce plc, Cranfield University, UK Atomic Energy Authority and MDA Space and Robotics. They acknowledge Professor Dominic O'Brien for the use of optical equipment. They thank Dr Richard Reeves, Prof Andong Wang, and Dr Stefan Kefer for their support with the equipment. They are also grateful to Dr Peter Walters, Dr Yue Liu, Dr Andy Schreier, and Dr Sheng Huang for the enlightening discussions.

**Disclosures.** The authors declare no conflicts of interest.

**Data availability.** Data underlying the results presented in this paper are available in Data File 1 and Data File 2 of the Supplementary Information.